\documentclass[12pt,showpacs,preprintnumbers,amsmath,amssymb]{article}
\usepackage[dvips]{graphicx}
\usepackage{amssymb}
\usepackage{amsmath}
\usepackage{graphicx,color}
\usepackage{amsfonts}
\usepackage{bm}
\usepackage{cancel}
\usepackage{comment}

\usepackage[colorlinks=true,linkcolor=black,unicode=true]{hyperref}
\usepackage[left=30mm, top=20mm, right=30mm, bottom=15mm, nohead, nofoot]{geometry}

\newcommand\be{\begin{equation}}
\newcommand\ee{\end{equation}}

\allowdisplaybreaks[4]

\begin{document}

\font\bss=cmr12 scaled
\magstep 0



\title{Compact Stars with Dark Energy in General Relativity and Modified Gravity}

\author{Artyom V. Astashenok$^{1}$, Sergey D. Odintsov$^{2,3}$, Vasilis K. Oikonomou,$^{4}$\\
\small $^{1}$Institute of Physics, Mathematics and IT, I.
Kant Baltic Federal University,\\ 
\small 236041 Kaliningrad, Russia\\
\small $^{2}$ICREA, Passeig Luis Companys, 23, 08010 Barcelona, Spain \\
\small $^{3}$Institute of Space Sciences (ICE,CSIC) C. Can Magrans s/n,
08193 Barcelona, Spain\\
\small $^{4}$ Department of Physics, Aristotle University of
Thessaloniki, Thessaloniki 54124, Greece
}
\date {}

\maketitle

\begin{abstract}
We investigate realistic models of compact objects, focusing on
neutron and strange stars, composed by dense matter and dark
energy in the form of a simple fluid or scalar field interacting
with matter. For the dark energy component, we use equations of
state compatible with cosmological observations. This requirement
 strongly constrains possible deviations from the simple $\Lambda$-Cold
Dark-Matter model with EoS $p_{d}=-\rho_{d}$ at least for small
densities of the dark component. But we can propose that the
density of dark energy interacting with matter can reach large
values in relativistic stars and affects the star parameters such
as the mass and radius. Simple models of dark energy are
considered. Then we investigated possible effects from modified
gravity choosing to study the $R^2$ model combined with dark
energy. Finally, the case of dark energy as scalar field
non-minimally interacting with gravity is considered.
\end{abstract}


\section{Introduction}

The physical nature of dark energy (DE) proposed as source of the
confirmed cosmic acceleration \cite{Riess1,Riess2,Perlmutter}
is one of the puzzles of modern physics. Very interesting theories
are considered in the literature (see \cite{Dark-6,Ca,Dark-7}).
According to observational data, baryon and dark matter compose
only the 28\% of the total energy budget of the Universe and the
remaining part is composed by DE \cite{Kowalski}. Concordance
with astrophysical observations requires negative pressure and
non-coupling with baryonic matter (or a very small coupling) for
DE.

Usually models of dark energy are investigated in cosmological
context. But it is interesting to consider possible manifestations
of dark energy in astrophysics of compact objects. One can assume
that for very high densities existing in the cores of neutron
stars, some interaction between dense matter and dark energy takes
place. Of course in this case we stepping on a rather shaky ground
of conjectures because we have no details for possible DE-matter
interaction. But assuming although general properties of such an
interaction, we can estimate possible influences of dark energy on
compact star properties.

Although the most successful model for the cosmological
acceleration is the $\Lambda$-Cold-Dark-Matter ($\Lambda$CDM)
model, it has several difficulties from the theoretical physics
viewpoint. One needs to answer difficult questions, such as why
the cosmological constant is very small. Large discrepancy between
observed value of $\Lambda$ term and its value predicted by any
quantum field theory, needs to be explained \cite{Weinberg}.
There is another way to explain cosmological acceleration in
frames of General Relativity (GR), namely it is the introduction
of a scalar field. From astrophysical observations it follows that
such field may be a phantom field with positive kinetic terms, and
therefore the dark energy equation of state (EoS) parameter can be
smaller than $-1$. Of course phantom fields are very problematic
from a quantum field theory point of view, but analysis of
observations do not exclude this scenario, in fact the latest
Planck data marginally allow a phantom dark energy possibility.

Another approach for resolution of the cosmic acceleration problem
is that acceleration is caused by deviations of gravity from
General Relativity (GR) theory on large cosmological distances.
The viable models of modified gravity are investigated in
literature
\cite{reviews1,reviews2,reviews4,book,reviews5,reviews6,dimo}.
In frames of modified gravity late- and early-time acceleration
can be described in a geometric way \cite{Nojiri:2003ft}.

Even not in the context of cosmological dynamics one can assume
that due to the strong gravitational fields inside compact stars,
nonlinear higher order curvature terms play some role. Therefore,
it is interesting to investigate the effects of the combination of
modified gravity and of a hypothetical interaction between dark
energy and matter.

Models of neutron stars in modified gravity are widely considered
in the literature  (see
\cite{Astashenok:2020qds,Astashenok:2021peo,Capozziello:2015yza,Astashenok:2014nua,Astashenok:2020cfv,Arapoglu:2010rz,Panotopoulos:2021sbf,Lobato:2020fxt,Oikonomou:2021iid,Odintsov:2021qbq},
and for a recent review see \cite{Wojnar3}). In particular for
the simplest $R^2$ gravity, the solution of the modified
Tolman-Oppenheimer-Volkoff equations asymptotically tends to
Schwarzschild solution on spatial infinity. The area near surface
of star gives some contribution to gravitational mass measured by
distant observer and in result the neutron star mass increases in
comparison with General Relativity for the same central density.
It allows to explain NS with large mass
\cite{Pani:2014jra,Doneva:2013qva,Horbatsch:2015bua,Silva:2014fca,Chew:2019lsa,Blazquez-Salcedo:2020ibb,Motahar:2017blm}.

In this paper we shall investigate realistic models of neutron and
strange stars, which are composed by dense matter and dark energy
in the form of a simple fluid or in the form of a scalar field
interacting with matter. For the dark energy component, we use
equations of state compatible with cosmological observations. This
requirement strongly constrains possible deviations from the
simple $\Lambda$CDM model with an EoS $p_{d}=-\rho_{d}$. Under the
assumption that dark energy interacts with dark matter, we shall
investigate the effects of this framework on the star parameters
such as the mass and radius. We also consider the combined effect
of the new theoretical framework we introduce with an $R^2$
gravity model.

The structure of this paper is as follows: In section 2 we briefly
consider Tolman-Oppenheimer-Volkoff equations in GR and dark
energy models in cosmology. We introduce an interaction between
dark energy and matter such that for only very high densities of
matter this interaction can influence the parameters of the
compact stars. With regard to the EoS of the dark energy
component, we choose a simple equation which slightly differs from
the simple equation of state for the cosmological constant, for
current density, but can lead to various dynamics apart from de
Sitter expansion. Then we calculate the parameters of the compact
stars for simple models of dark energy and for several well-known
EoS for stellar matter. Our calculations demonstrate that the
mass-radius relation and the dependence of mass from the central
density changes in comparison with results in General Relativity
in a manner which is not depended from the EoS for stellar matter.
In section 4 compact stars models are considered in conjunction
with the effects of $R^2$ gravity alongside with possible effects
from dark energy interacting with matter. The last section is
devoted to dark energy considered as a scalar field non-minimally
interacting with the gravitational field. The conclusions follow
in the end of the paper.

\section{Tolman-Oppenheimer-Volkoff equations and models of dark energy}

We start from the Einstein equations in GR, namely,
\begin{equation}
    R_{\mu\nu}-\frac{1}{2}g_{\mu\nu}R = 8\pi T_{\mu\nu},\label{EinEq}
\end{equation}
where $R_{\mu\nu}$ is the Ricci tensor, $R$ denotes the Ricci
scalar, and $T_{\mu\nu}$ is the energy-momentum tensor. Hereafter,
we shall consider a physical units system in which $G=c^2=1$. For
non-rotating isotropic stars we assume a spherically symmetric
metric with line element,
\begin{equation}
    ds^2 = -e^{2\nu}dt^2 + e^{2\lambda} dr^2 + r^2 d\Omega^2.\label{metr_star}
\end{equation}
In the simplest case that the energy-momentum tensor is composed
of matter and dark energy, for ideal fluids we have,
\begin{equation}
    T_{0}^{0} = - (\rho_{m} + \rho_{d}),\quad  T_{i}^{i} =   p_{m} + p_{d}, \quad T_{0}^{i} = 0.
\end{equation}
With this assumption one can obtain following equations for metric functions:
\begin{equation}
e^{-2\lambda}\left[\frac{2\lambda'}{r}-\frac{1}{r^{2}} \right]+\frac{1}{r^{2}}=8\pi(\rho_{m}+\rho_{d}),   \label{fe1}
\end{equation}
\begin{equation}
e^{-2\lambda}\left[\frac{1}{r^{2}}+\frac{2\nu'}{r} \right]-\frac{1}{r^{2}}=8\pi (p_{m}+p_{d}), \label{fe2}
\end{equation}
\begin{equation}
e^{-2\lambda}\left[ \nu'^{2}+\nu''-\lambda'\nu'+\frac{1}{r}(\nu'-\lambda')\right]=8\pi (p_{m}+p_{d}).\label{fe3}
\end{equation}
where the $()^{\prime}$ denotes derivative radial coordinate.
Combining Eqs. (\ref{fe2}), (\ref{fe3}) one obtains the equation,
\begin{equation}
    \frac{dp}{dr} = -(\rho + p)\frac{d\nu}{dr}, \quad p = p_m + p_d, \quad \rho = \rho_m + \rho_d. \label{fe4}
\end{equation}
The next step is to define how dark energy interacts with matter.
Unfortunately we have no indications on this from observational
data. Firstly one notes that for small densities there is no
coupling between dark energy and matter. Obviously this assumption
can be extended from cosmological densities to densities of
ordinary stars and even of white dwarfs because these
astrophysical objects are well described in the context of GR,
without any additional assumptions. For our illustrative
calculations we consider the following coupling between matter and
dark energy,
\begin{equation}
    \rho_d = \alpha \rho_m \exp(-\rho_s / \rho_m),
\end{equation}
where $\alpha$ is a positive constant. The exponential factor
leads to sharp decreasing of the corresponding dark energy density
for $\rho_m<<\rho_s$ where $\rho_s$ is some cut-off factor. For
our calculations we choose $\rho_s = 300$ MeV/fm$^3$.

The next step is to define the EoS for dark energy. It is
interesting to consider models of dark energy corresponding on
cosmological level to various cosmological dynamics. Let us
briefly discuss cosmological aspects of dark energy. The Friedmann
equations for a spatially flat universe with metric
\begin{equation}
ds^2_{univ} = - dt^2 + a^2(t)(dx^2 + dy^2 + dz^2)\label{metr_univ}
\end{equation}
are
\begin{equation}
    \frac{\dot{a}^2}{a^2} = \frac{8\pi}{3}\rho,\label{fr1}
\end{equation}
\begin{equation}
    \dot{\rho} + 3\frac{\dot{a}}{a}\left(\rho + p\right)=0.\label{fr2}
\end{equation}
Here $a$ is the scale factor and the ``dot'' indicates the
derivative with respect to the cosmic time. The last equation is
satisfied for any component. For the dark energy fluid it is
convenient to use an EoS of the form,
\begin{equation}
    p_d = -\rho_d + f(\rho_{d})\label{EoS}
\end{equation}
Standard cosmological model states that the Universe is filled
radiation, cold dark matter and non-zero vacuum energy or
cosmological constant with simple relation between energy density
and pressure $p_{\Lambda} = -\rho_{\Lambda}$.

For the current $\rho_d$ function we assume that
$|f(\rho_d)|<<\rho_d$ because we know that the $\Lambda$CDM model
fits well the observational data and therefore possible deviations
from it in any case might be negligible. But future cosmological
evolution of the Universe can differ from the pure de Sitter
expansion very strongly even if $f(\rho_d)<<\rho_d$  at present
time.

Neglecting matter and radiation one can obtain from the equation
of the Hubble parameter the following relation for time $t$ from
present epoch $t_0$,
\begin{equation}
    t - t_{0} = \frac{1}{\sqrt{24\pi}}\int_{\rho_{d0}}^{\rho_d}\frac{d\rho_d}{\rho_d^{1/2} f(\rho_d)}.\label{trho}
\end{equation}
For quintessence, the energy density decreases with time ($f>0$,
$\rho_d<\rho_{d0}$) while as for phantom field ($f<0$) dark energy
density increases. From the continuity equation we have for the
scale factor as function of energy density,
\begin{equation}
a = a_{0}\exp\left(\frac{1}{3}\int_{\rho_{d0}}^{\rho_d}\frac{d\rho_d}{f(\rho_d)}\right).\label{arho}
\end{equation}
For example if we consider $f(\rho_d)$ in the form,
\begin{equation}
f(\rho_d) =  \beta\rho_{d}^{m}, \quad m>0.\label{EoS_1}
\end{equation}
Positive values of $\beta$ correspond to quintessence and negative
to phantom energy. For the phantom case, two ways of evolution in
the future are possible. If $0<m\leq 1/2$ the energy density
approaches an infinite value at $t\rightarrow\infty$. The
so-called Little Rip scenario takes place. A finite time
singularity occurs for $m>1/2$ because the integral for time
converges at $\rho_d\rightarrow\infty$. The scale factor diverges
($1/2\leq m \leq 1$, Big Rip singularity) or even remains finite
($m>1$, singularity of type III according to classification of
\cite{NOT}).

We can consider also a class of dark energy models with asymptotic
de Sitter evolution. It is realized for example if dark energy EoS
has the form \be\label{EOSDS} f(\rho_{d})=-\beta
\rho_{f}\left(1-\sqrt{\frac{\rho_{d}}{\rho_{f}}}\right), \quad
0<\rho_{d}<\rho_{f} \mbox{ or }  \rho_{d}>\rho_{f}. \ee Here the
dimensionless positive parameter $\beta>0$ is introduced. The dark
energy density asymptotically tends to $\rho_{f}$ from below
(phantom energy) or above and the Universe expands according to de
Sitter law at $t\rightarrow\infty$ (so called Pseudo-Rip).

Obviously however at an astrophysical level, such models of dark
energy differ very negligibly from a model with
$p_d\approx-\rho_d$. For the applicability of this new framework,
one needs to assume that $\rho_f>>\rho_{d0}$. This requirement in
its turn means that $\beta<<\rho_{d0}$. Therefore,
$|f(\rho_d)|<<\rho_d$ for typical densities in compact stars.
Otherwise in a reliable model with (\ref{EoS_1}) for large
densities $|f(\rho_d)|$ can considerably exceed $\rho_d$ and such
model does not satisfy the astronomical constraints coming from
observations. From this perspective, we can conclude that for a
wide class of reliable dark energy models the EoS is $p_d\approx
-\rho_d$ not only for densities on cosmological scales, but for
relativistic densities also. Below we consider in detail this case
and a model with  the EoS (\ref{EoS_1}) as example of model for
which pressure can differ considerably from $-\rho_d$ for
$\rho_d>>|\beta|^{\frac{1}{1-m}}$.

\section{Mass-radius and mass-density relation for compact stars in GR with dark energy}

\textbf{EoS for relativistic stars}.

We investigate possible influences of dark energy on parameters of
neutron stars, and hypothetical quark stars. For describing matter
in these relativistic objects, one needs to use EoSs from particle
physics.

For neutron stars many EoS based on various approaches are
proposed. We use the well-known AP4 EoS \cite{AP4} derived
from many body calculations with  three-nucleon potential and
Argonne 18 potential with UIX potential. As an example of an EoS
obtained from relativistic Dirac-Brueckner-Hartree-Fock formalism,
the MPA1 EoS (\cite{MPA}) is considered. Finally, we include in
our consideration the GM1 EoS based on relativistic mean-field
theory. This EoS was proposed by \cite{GM} and then extended in
\cite{GMnph} for cold neutron matter containing the baryon
octet.

As assumed quark (or so-called strange) stars consist from
deconfined quarks u, d, s and electrons (see \cite{QS},
\cite{QS1}). Because deconfined quarks form color
superconductor system, the EoS became softer in comparison with
the standard hadron matter. Possible observable effects include
minimum allowed mass, radii, cooling behavior and other.

The EoS for strange matter is very simple in the so-called MIT bag model namely \cite{MIT}, \cite{MIT1}:
\begin{equation}
\label{QEOS} p=b(\rho-4B),
\end{equation}
where $B$ is the bag constant. The value of the parameter $b$
depends on both the chosen mass for the strange quark $m_s$ and
QCD coupling constant and usually varies from $b=1/3$ ($m_s=0$) to
$b=0.28$ ($m_{s}=250$ MeV). The value of $B$ lies in interval
$0.98<B<1.52$ in units of $B_{0}=60$ MeV/fm$^{3}$
\cite{MIT2}. In the following, we shall consider the case
$B=1$ and $b=0.31$.

\textbf{Simple model of dark energy $\rho_d \approx -p_d$}.

We start our considerations using the simplest model of dark
energy, namely when $f_{d}\approx 0$ (effective cosmological
constant). Obviously, negative contribution of dark energy
pressure leads to effective softening of nuclear EoS and mass of
stellar configuration decreases in comparison with GR for given
central density. For illustrative calculations we choose two
values of dimensionless parameter $\alpha$, $0.025$ and $0.050$.
From our calculations it follows (see figures 1-3) that the total
decrease of the maximal mass and the corresponding radius depends
on the parameter $\alpha$ in a linear way. For the APR4, GM1 and
MPA1 EoS, increasing of $\alpha$ at $0.025$ corresponds to a
decrease of maximal mass of the order $0.07-0.09$ $M_\odot$ and
the corresponding decrease of radii by $0.3-0.4$ km. For the GM1
EoS, the mass and radius decrease more. Calculations for strange
stars leads to same results.

The dependence of the mass from radius for various values of
$\alpha$ is shifted, in comparison with $M-R$ relation for neutron
stars without hypothetical dark energy. One can see that this
shift is close to be uniform for masses greater than $0.5
M_{\odot}$.
\begin{figure}
    \centering
    \includegraphics[scale=0.35]{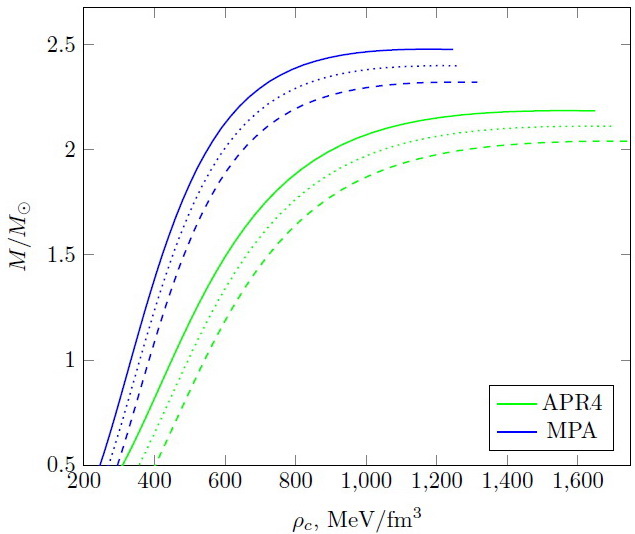}\includegraphics[scale=0.35]{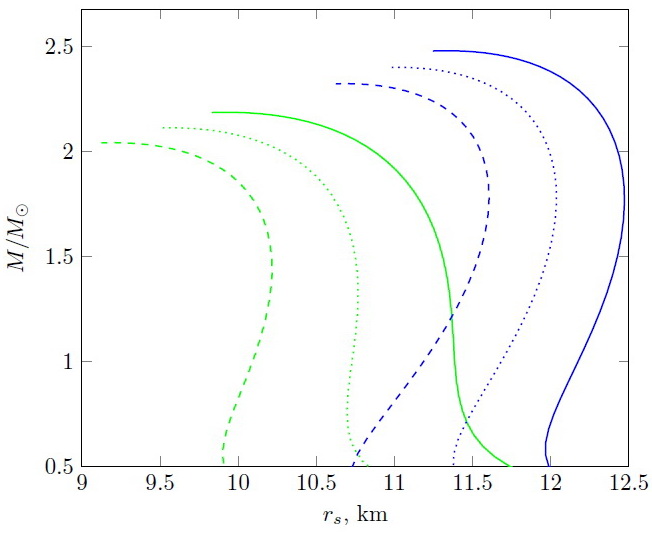}
    \caption{The dependence of stellar mass from central density (left panel)
and mass-radius relation (right panel) for the APR4 and MPA1 EoS in
a case of simple model of coupling between matter and dark energy
for $p_d\approx - \rho_d$. Solid, dotted and dashed lines
correspond to $\alpha=0$, $0.025$ and $0.05$. For cut-off density
$\rho_s$ we choose 300 MeV/fm$^3$.}
    \label{fig1}
\end{figure}

\begin{figure}
    \centering
    \includegraphics[scale=0.35]{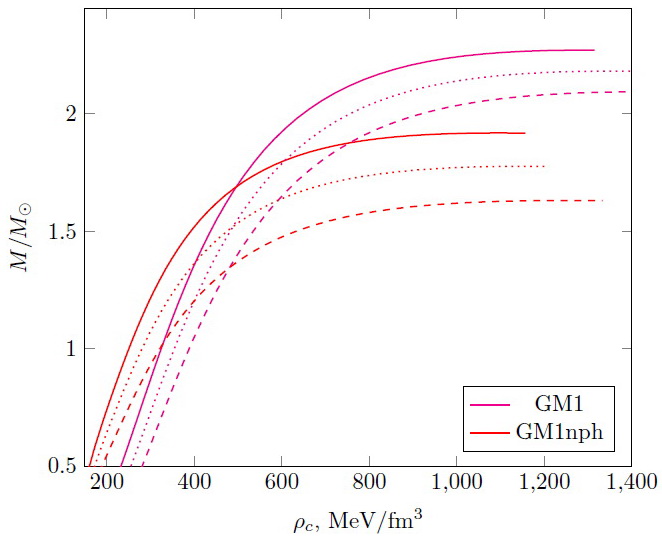}\includegraphics[scale=0.35]{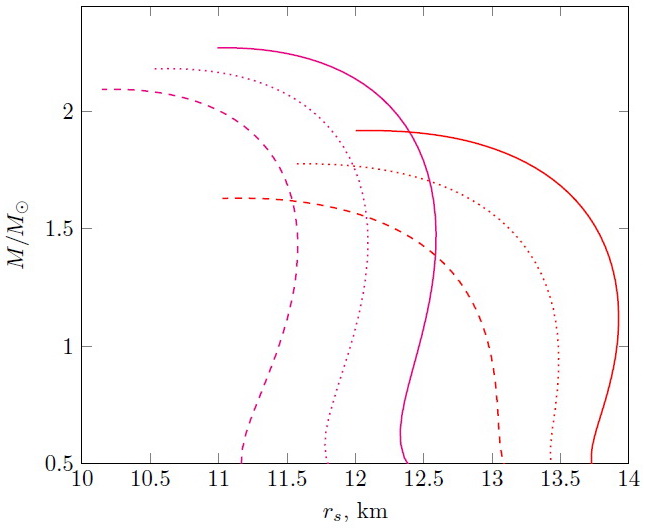}
    \caption{The same as in Fig. 1 but for the GM1 EoS and its extension for hyperon sector GM1 EoS.}
    \label{fig2}
\end{figure}

\begin{figure}
    \centering
    \includegraphics[scale=0.35]{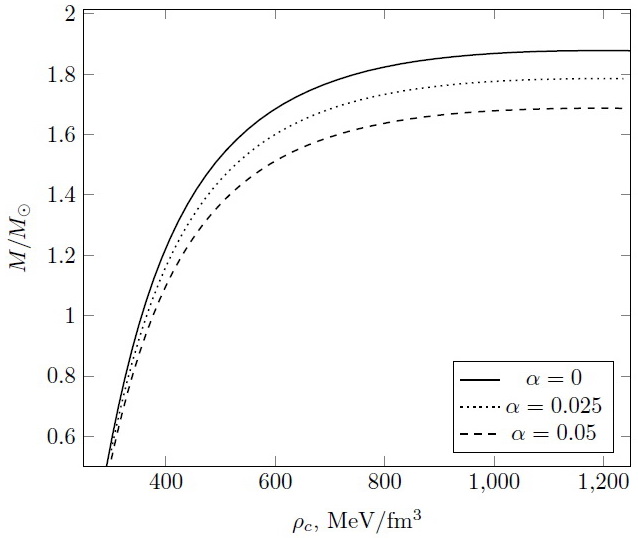}\includegraphics[scale=0.35]{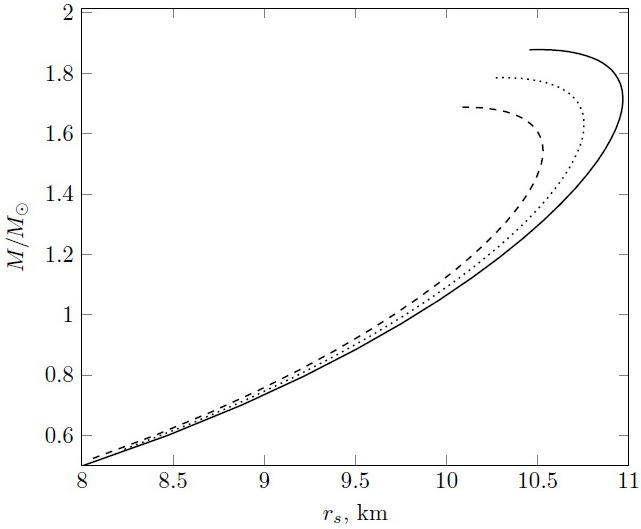}
    \caption{The same as on Fig. 1 but for strange stars.}
    \label{fig3}
\end{figure}
From our calculations it follows that the value $\alpha=0.05$ is
critical for the GM1 and APR4 EoSs because the maximal mass for
this $\alpha$ is close to the upper limit of the mass (2.05
$M_{\odot}$) of slow-rotating pulsar PSR J0348+0432
\cite{Antoniadis}, \cite{Demorest}. One needs to mention
also the heaviest black widow pulsar J0952-0607 \cite{Romani}
with a mass $2.35^{+0.17}_{-0.17} M_{\odot}$ but its rotation
frequency is nearly $700$ Hz and therefore effects from fast
rotation can significantly increase mass in comparison with
non-rotating stellar configuration.

In light of these data, EoSs with hyperons (GM1) can be considered
as unrealistic as they are in GR not only in our simplest model of
dark energy. We investigated this EoS for illustrative purposes.
For soft EoSs, the effects of dark energy on the neutron star mass
is more stronger than in a case a stiff EoS is used.

For quark stars the dependence of the radius from the mass
significantly changes only for large masses close to maximal. The
maximal difference between radii for stars with the same mass is
only around $\sim $ km (for $\alpha=0.05$) while for neutron stars
this discrepancy can exceeds $2$ km (for soft EoS GM1 is $>2.5$
km).

The obtained results are intuitively clear. The addition of dark
energy with EoS parameter $w=-1$ leads to softening of effective
EoS for matter and the corresponding decreasing of mass and radius
for a given density of matter at the center.

Then we consider another model with EoS (\ref{EoS_1}) for $m=2$.
From previous issue one can conclude that for negative values of
$\beta$, the effect of EoS softening will intensify even more, and
most of the equations of state in such model would not satisfy the
observed limit of neutron star mass. Another possibility is that
$\beta^{-1}>>\rho_m$  and this model is equivalent to the one
considered above.

Therefore we choose positive values of $\beta$, varying from 0 to
$1$ in units of $\beta_0 = $(5 MeV/fm$^3$)$^{-1}$. At a
cosmological level, this model of course is indistinguishable from
a pure cosmological constant, because the current dark energy
density $\rho_{d0}$ is very negligible in comparison with the
chosen value of $\beta^{-1}$.
\begin{figure}
    \centering
    \includegraphics[scale=0.35]{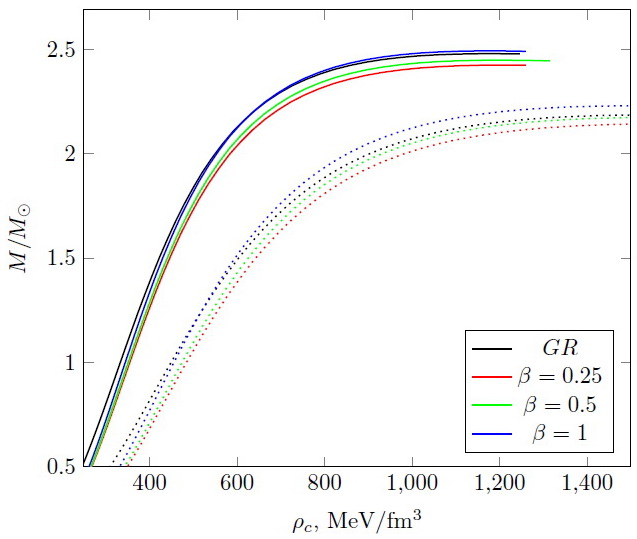}\includegraphics[scale=0.35]{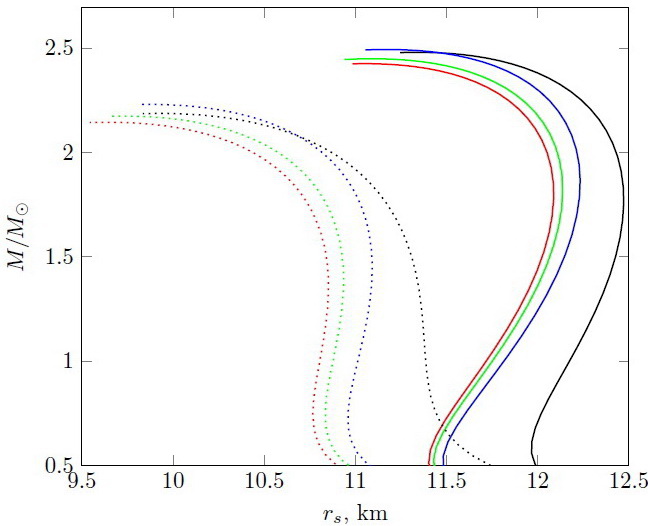}
    \caption{The dependence of stellar mass from the central density (left panel)
and mass-radius relation (right panel) for MPA1 (solid lines) and
APR4 (dotted lines) EoSs in the case of a  simple model of
coupling between matter and dark energy with $p_d = - \rho_d +
\beta\rho_d^{2}$ for various values of $\beta$ in units of (5
MeV/fm$^{3}$)$^{-1}$.}
    \label{fig4}
\end{figure}

\begin{figure}
    \centering
    \includegraphics[scale=0.33]{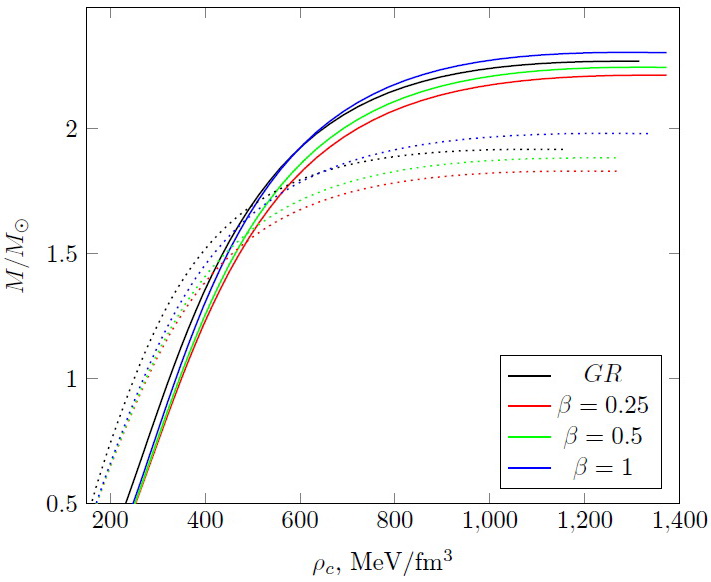}\includegraphics[scale=0.33]{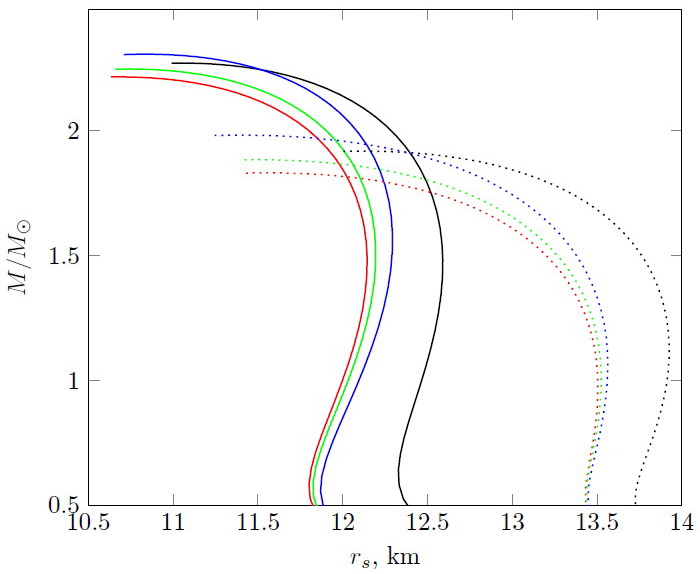}
    \caption{The same as in Fig. 4 but for GM1 (solid lines) and GM1nph (dotted lines) EOS.}
    \label{fig5}
\end{figure}

\begin{figure}
    \centering
    \includegraphics[scale=0.35]{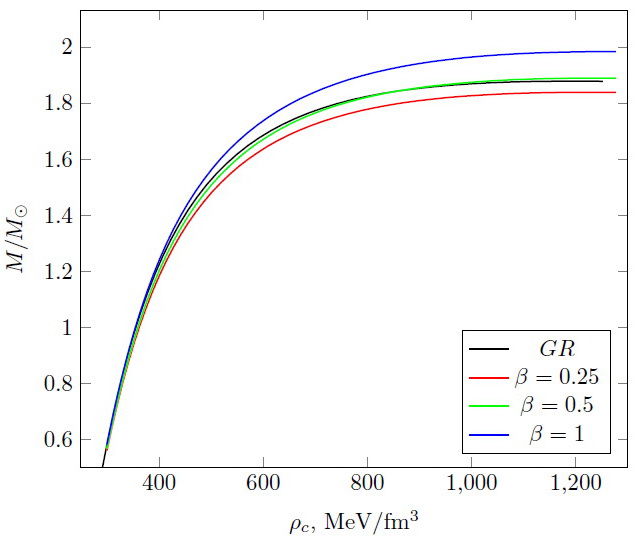}\includegraphics[scale=0.35]{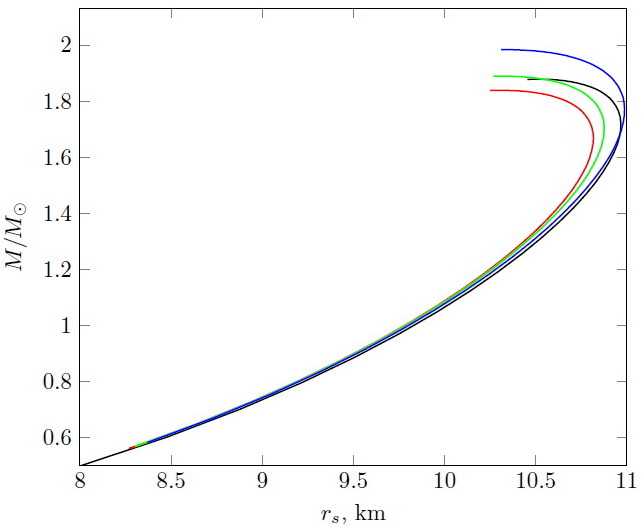}
    \caption{The same as on Fig. 4 but for quark stars.}
    \label{fig6}
\end{figure}
In table 1 we present the parameters of neutron stars for various
EoSs and some values of $\alpha$ and $\beta$. Also results are
presented on Fig. 4-6. The maximal mass of the star increases with
increasing of the parameter $\beta$, because this corresponds to
positive contribution to total pressure. In the case that
$\alpha=0.025$ for $\beta\sim 1$, the maximal mass slightly
increases in comparison with GR for hadronic EoSs while the radius
of the stellar configuration decreases. For GM1nph EoS this effect
is stronger.
\begin{table}
\begin{center}
\begin{tabular}{c|c|c|c|c|c}
\hline
$\alpha$ & $\beta,$ & $M_{max},$ & $R_{max},$ & $R_{1.4},$ & $R_{1.0},$ \\
         & $\beta_{0}$ & $M_{\odot}$ & km & km & km \\ \hline
\multicolumn{6}{c}{APR4}\\ \hline
    0 &  0 & 2.19 & 9.91 & 11.33 & 11.39\\ \hline
    0.025 & 0 &  2.11 & 9.57 & 10.76 & 10.73 \\ \hline
    0.050 & 0 &  2.04 & 9.19 & 10.21 & 10.09 \\ \hline
    0.025 & 0.25 & 2.14 & 9.61 & 10.85 & 10.80 \\ \hline
    0.025 & 0.50 & 2.17 & 9.73 & 10.93 & 10.87 \\ \hline
    0.025 & 1   & 2.23 & 9.89 & 11.09 & 11.00 \\ \hline
    0.05 & 0.25 & 2.16 & 9.53 & 10.56 & 10.40 \\ \hline
    0.05 & 0.50 & 2.26 & 9.78 & 10.86 & 10.67 \\ \hline

\multicolumn{6}{c}{MPA}\\ \hline
    0 &  0 & 2.48 & 11.31 & 12.39 & 12.17 \\ \hline
    0.025 & 0 &  2.40 & 11.03 & 11.94 & 11.66 \\ \hline
    0.050 & 0 & 2.32 & 10.69 & 11.49 & 11.18  \\ \hline
    0.025 & 0.25 &  2.43 & 11.05 & 11.98 & 11.72 \\ \hline
    0.025 & 0.50 &  2.45 & 11.00 & 12.02 & 11.76 \\ \hline
    0.025 & 1 &  2.49 & 11.13 & 12.10 & 11.80 \\ \hline
    0.050 & 0.25 & 2.42 & 10.87 & 11.66 & 11.34  \\ \hline
    0.050 & 0.50 & 2.50 & 10.97 & 11.81 & 11.47  \\ \hline

\multicolumn{6}{c}{GM1}\\ \hline
    0 &  0 & 2.27 & 11.07 & 12.59 & 12.46 \\ \hline
    0.025 &  0 & 2.18 & 10.61 & 12.09 & 11.96 \\ \hline
    0.050 &  0 & 2.09 & 10.22 & 11.58 & 11.44 \\ \hline
    0.025 &  0.25 & 2.22 & 10.72 & 12.14 & 12.00 \\ \hline
    0.025 &  0.50 & 2.25 & 10.74 & 12.18 & 12.03 \\ \hline
    0.025 &  1 & 2.31 & 10.79 & 12.27 & 12.09 \\ \hline
    0.05 &  0.25 & 2.23 & 10.42 & 11.79 & 11.59 \\ \hline
    0.05 &  0.50 & 2.33 & 10.72 & 11.98 & 11.73 \\ \hline

\multicolumn{6}{c}{GM1nph}\\ \hline
    0 &  0 & 1.92 & 12.08 & 13.84 & 13.91\\ \hline
    0.025 &  0 & 1.78 & 11.66 & 13.27 & 13.48 \\ \hline
    0.05 &  0 & 1.63 & 11.11 & 12.54 & 12.98 \\ \hline
    0.025 &  0.25 & 1.83 & 11.53 & 13.32 & 13.50 \\ \hline
    0.025 &  0.50 & 1.88 & 11.49 & 13.37 & 13.52 \\ \hline
    0.025 &  1 & 1.98 & 11.34 & 13.46 & 13.56 \\ \hline
    0.05 &  0.25 & 1.86 & 10.78 & 12.88 & 13.09 \\ \hline
    0.05 &  0.50 & 2.05 & 10.89 & 13.08 & 13.19 \\ \hline

\end{tabular}
\caption{The parameters of neutron stars for various EoSs and
various values of $\alpha$ and $\beta$ in a model of dark energy
with $p_d = -\rho_d + \beta \rho_{d}^{2}$ and coupling with matter
as $\rho_d = \alpha \rho_m$. The parameter $\beta$ is given in
units of $\beta_{0}=$ (5 MeV/fm$^3$)$^{-1}$. $R_{1.4}$ and
$R_{1.0}$ mean radii of star with masses 1.4 and 1.0 $M_{\odot}$
correspondingly.}
\end{center}
\end{table}

\section{Spherically symmetric stellar configurations in R-square gravity}

Now investigate possible effects from corrections to GR due to
large values of spatial curvature. To account these corrections at
least at first order approximation ($R^2$ gravity) one needs to
replace the standard Einstein-Hilbert action for the gravitational
field which contains the scalar curvature $R$ by the following
way:
\begin{equation}\label{action}
S_g=\frac{1}{16\pi}\int d^4x \sqrt{-g}\left(R+\gamma R^2\right).
\end{equation}
Varying the action with respect to  $g_{\mu\nu}$, gives us the
equation of motion for metric functions:
\begin{equation}\label{field}
(1+2\gamma R)G_{\mu \nu }+\frac{1}{2}\gamma R^2 g_{\mu \nu }-2(\nabla
_{\mu }\nabla _{\nu }-g_{\mu \nu }\Box )R=8 \pi T_{\mu \nu }.
\end{equation}
Here $G_{\mu\nu}=R_{\mu\nu}-\frac{1}{2}Rg_{\mu\nu}$ is the
Einstein tensor.

The components of the field equations is nothing else than
Tolman-Oppenheimer-Volkoff equations in the frame of modified
gravity:
 \be \label{TOV1} \frac{1+2 \gamma R}{r^2}\left
[r\left(1-e^{-2\lambda }\right)\right]'=8\pi
\rho+\frac{\alpha}{2}R^2+ \ee
$$
+2\gamma e^{-2\lambda}\left[\left(\frac{2}{r}-{\lambda}'\right)R'+R''\right]
$$

\be \label{TOV2} \frac{1+2\gamma R}{r}
\left[2e^{-2\lambda}{\nu}'-\frac{1}{r}\left(1-e^{-2\lambda}\right)\right]
=8\pi
p+\frac{\gamma}{2}R^2+\ee
$$
+2\gamma e^{-2\lambda}\left(\frac{2}{r}+{\nu}'\right)R'
$$

\begin{figure}
    \centering
    \includegraphics[scale=0.35]{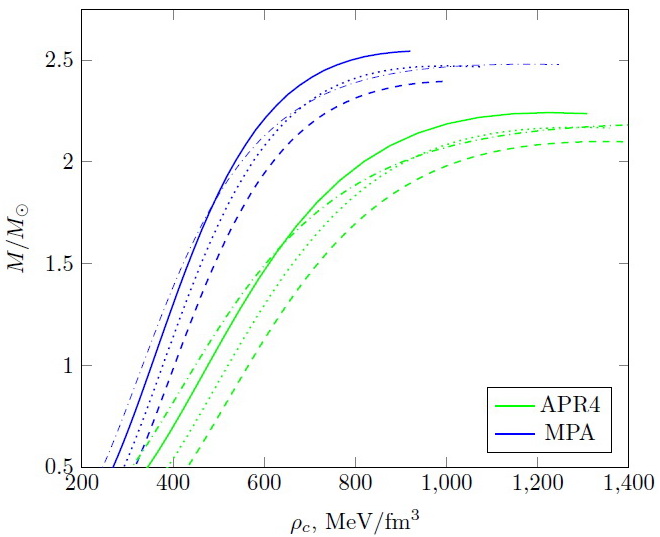}\includegraphics[scale=0.35]{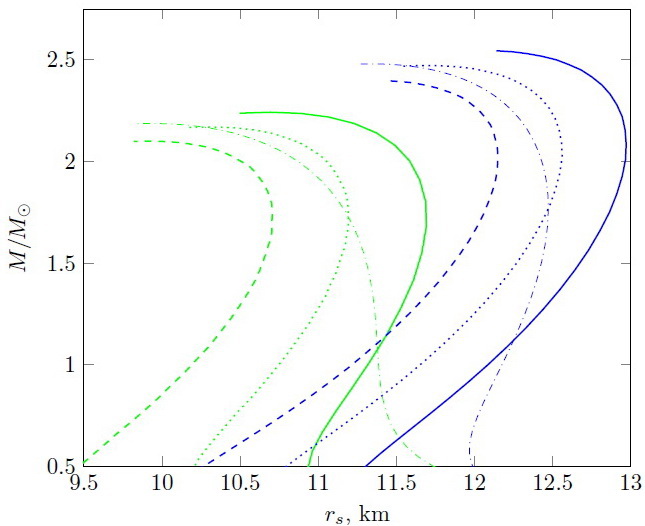}
    \caption{The dependence of stellar mass as a function of the central density (left panel)
and mass-radius relation (right panel) in $R^2$ gravity (parameter
$\gamma=2.5$ in units of $r_{g}^{2}=2GM_{\odot}/c^2$) for APR4 and
MPA1 EoS in a case of simple model of coupling between matter and
dark energy for $p_d\approx - \rho_d$. Solid, dotted and dashed
lines correspond to $\alpha=0$, $\alpha=0.025$ and $\alpha=0.05$.
Dash dotted lines correspond to GR.}
    \label{fig7}
\end{figure}

\begin{figure}
    \centering
    \includegraphics[scale=0.35]{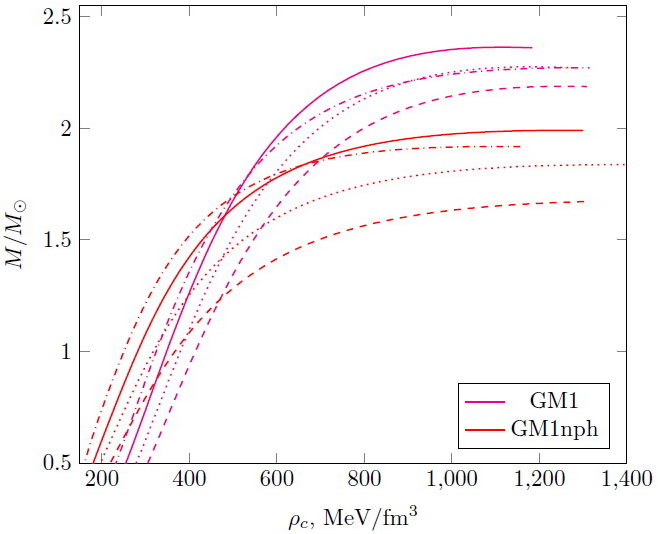}\includegraphics[scale=0.35]{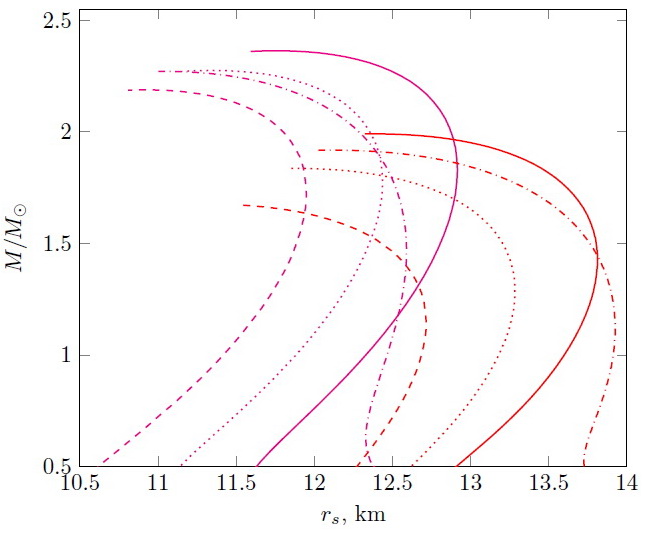}
    \caption{The same as in Fig. 7 but for GM1 and GM1nph EoS.}
    \label{fig8}
\end{figure}

\begin{figure}
    \centering
    \includegraphics[scale=0.35]{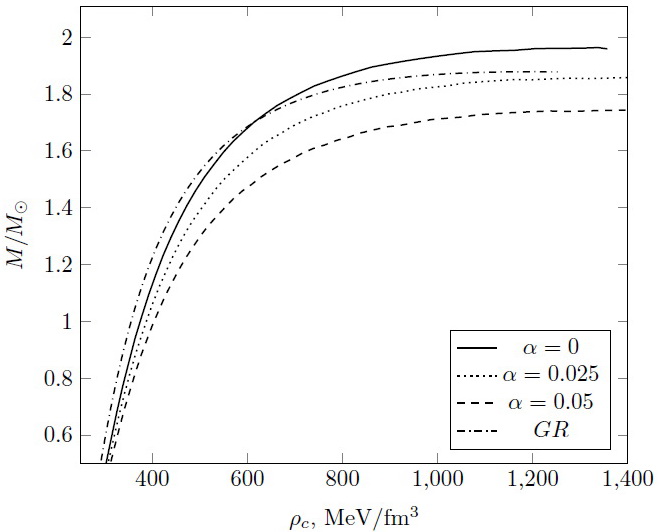}\includegraphics[scale=0.35]{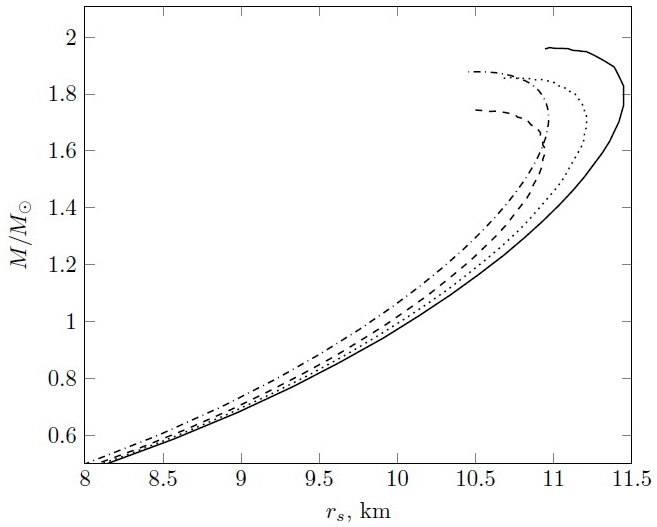}
    \caption{The same as in Fig. 7 but for strange stars.}
    \label{fig9}
\end{figure}
Our analysis indicates that for these equations, one can construct
solutions such that on spatial infinity it asymptotically
approaches the Schwarzschild solution. One needs to impose
following conditions on the value of the scalar curvature and its
derivative on $r$:
$$
R\rightarrow 0, \quad R'\rightarrow 0\quad \mbox{for} \quad r\rightarrow\infty.
$$
The gravitational mass for the observer far from the star can be
calculated in the infinite radius limit,
$$
M = \lim_{r\rightarrow\infty}\frac{r}{2}\left(1-e^{-2\lambda}\right).
$$
The results of our calculations for simple $\Lambda$-like model of
dark energy model are presented in Figs. 7-9. The effect of
increasing the mass due to $R$-square terms is partly eliminated
by the dark energy component for $\alpha\approx 0.025$ (for large
masses mass-radius relation is close to such in GR). The
mass-radius relation for neutron stars looks like that of strange
stars up to small masses (radius decreases with mass). Another
interesting feature is that radii for neutron stars with small
masses are considerably reduced in comparison with GR. For
considered parameters maximal difference for $M\approx 0.5
M_{\odot}$ can exceed 1.5 km (MPA1 EoS) or even 2 km (APR4 EoS).

For more interesting models of dark energy with $p_d = -\rho_d +
\beta\rho_d^2$, the effects from possible dark energy component
and $R-$square term reinforce each other (see Figs. 10-12).
\begin{figure}
    \centering
    \includegraphics[scale=0.42]{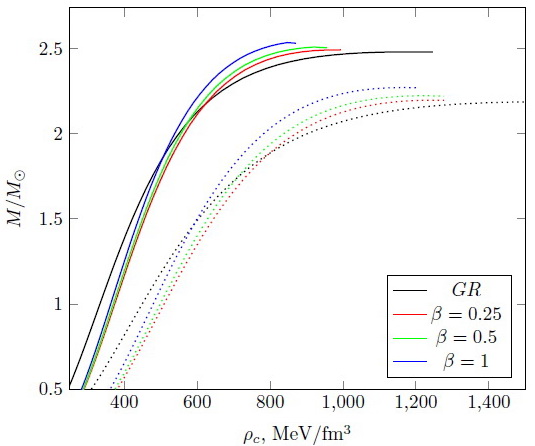}\includegraphics[scale=0.42]{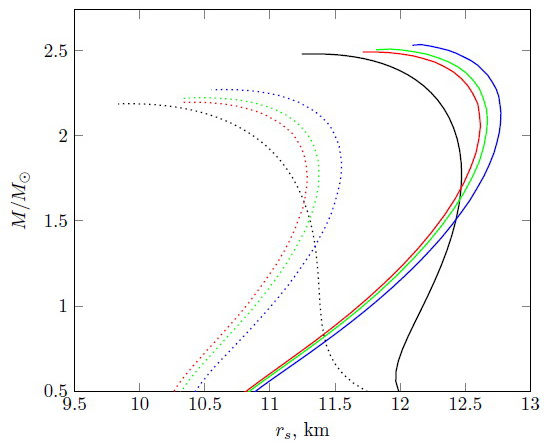}
    \caption{The dependence of the stellar mass from the central density (left panel)
and mass-radius relation (right) in $R^2$ gravity (parameter
$\gamma=10$ in units of $GM_{\odot}/c^2$) for APR4 (dotted lines)
and MPA1 EoS (solid lines) in a case model of coupling between
matter and dark energy for $p_d = - \rho_d + \beta\rho_{d}^{2}$.
The parameter $\alpha=0.025$ and the parameter $\beta$ is given in
units of $\beta_{0} = (5$ MeV/fm$^3$)$^{-1}$. Black lines
correspond to GR.}
    \label{fig10}
\end{figure}

\begin{figure}
    \centering
    \includegraphics[scale=0.35]{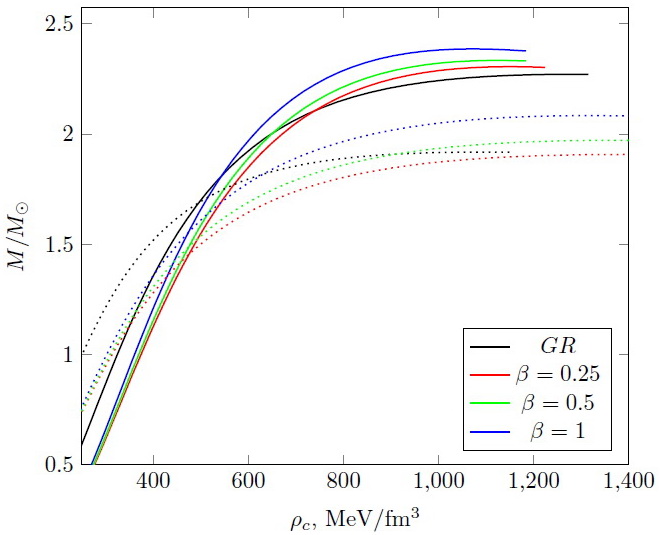}\includegraphics[scale=0.35]{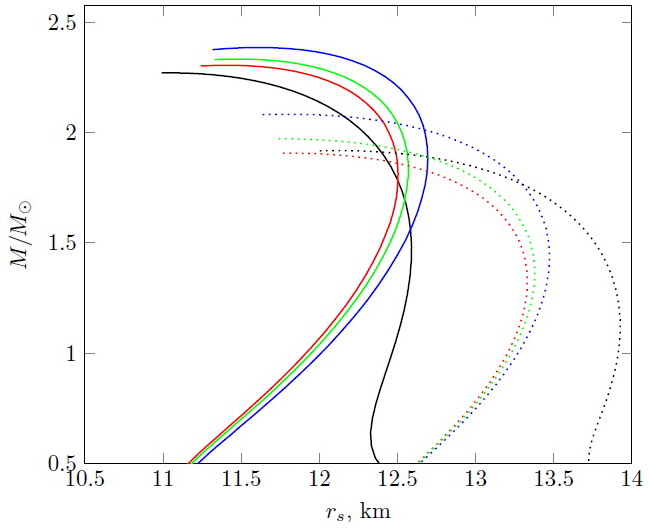}
    \caption{The same as in Fig. 10 but for GM1 (solid lines) and GM1nph (dotted lines) EoS.}
    \label{fig11}
\end{figure}

\begin{figure}
    \centering
    \includegraphics[scale=0.35]{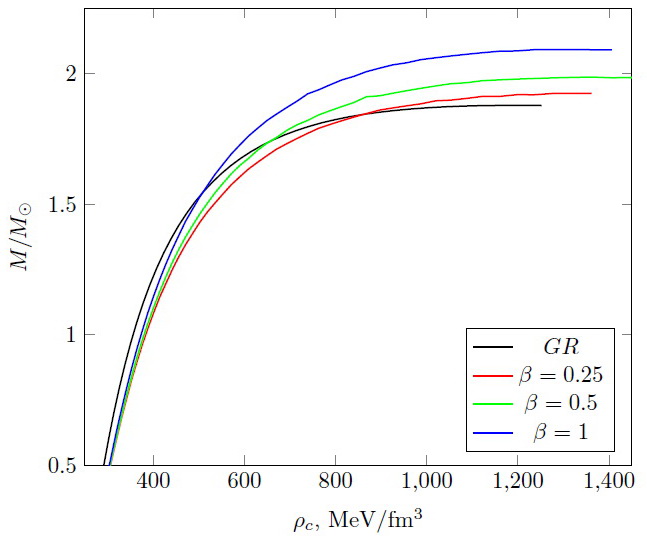}\includegraphics[scale=0.35]{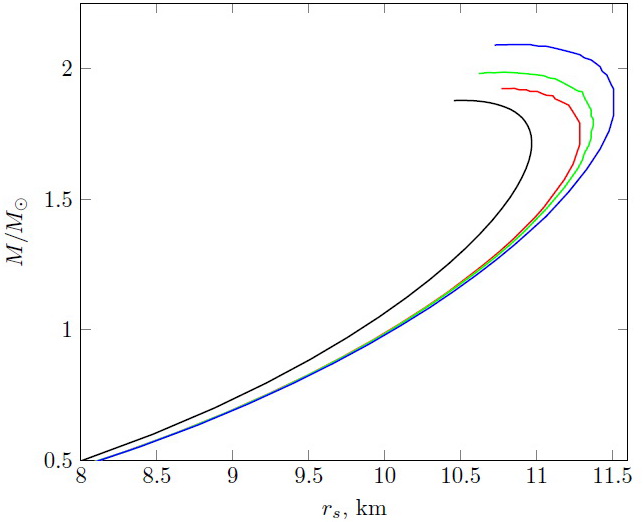}
    \caption{The same as in Fig. 10 but for strange stars.}
    \label{fig12}
\end{figure}
Maximal masses, corresponding to radii of neutron stars with
masses $M_{\odot}$ and $1.4M_{\odot}$ for $\gamma=10$ in units of
$GM_{\odot}/c^2$ are given in table 2.
\begin{table}
\begin{center}
\begin{tabular}{c|c|c|c|c|c}
\hline
$\alpha$ & $\beta,$ & $M_{max},$ & $R_{max},$ & $R_{1.4},$ & $R_{1.0},$ \\
         & $\beta_{0}$ & $M_{\odot}$ & km & km & km \\ \hline
\multicolumn{6}{c}{APR4}\\ \hline
    0 &  0 & 2.24 & 10.69 & 11.61 & 11.33\\ \hline
    0.025 & 0 &  2.17 & 10.23 & 11.09 & 10.75 \\ \hline
    0.050 & 0 &  2.10 & 9.90 & 10.59 & 10.18 \\ \hline
    0.025 & 0.25 & 2.20 & 10.42 & 11.16 & 10.82 \\ \hline
    0.025 & 0.50 & 2.22 & 10.43 & 11.25 & 10.88 \\ \hline
    0.025 & 1   & 2.27 & 10.65 & 11.38 & 11.00 \\ \hline
    0.05 & 0.25 & 2.20 & 10.31 & 10.91 & 10.50 \\ \hline
    0.05 & 0.50 & 2.29 & 10.55 & 11.18 & 10.72 \\ \hline

\multicolumn{6}{c}{MPA}\\ \hline
    0 &  0 & 2.54 & 12.14 & 12.59 & 12.08 \\ \hline
    0.025 & 0 &  2.47 & 11.61 & 12.17 & 11.67 \\ \hline
    0.050 & 0 & 2.39 & 11.52 & 11.77 & 11.20  \\ \hline
    0.025 & 0.25 &  2.49 & 11.81 & 12.22 & 11.68 \\ \hline
    0.025 & 0.50 &  2.51 & 11.92 & 12.25 & 11.72 \\ \hline
    0.025 & 1 &  2.53 & 12.16 & 12.32 & 11.75 \\ \hline
    0.050 & 0.25 & 2.48 & 11.64 & 11.92 & 11.36  \\ \hline
    0.050 & 0.50 & 2.53 & 12.04 & 12.08 & 11.47  \\ \hline

\multicolumn{6}{c}{GM1}\\ \hline
    0 &  0 & 2.36 & 11.68 & 12.74 & 12.33 \\ \hline
    0.025 &  0 & 2.27 & 11.25 & 12.29 & 11.88 \\ \hline
    0.050 &  0 & 2.19 & 10.89 & 11.83 & 11.42 \\ \hline
    0.025 &  0.25 & 2.30 & 11.33 & 12.34 & 11.92 \\ \hline
    0.025 &  0.50 & 2.33 & 11.42 & 12.39 & 11.96 \\ \hline
    0.025 &  1 & 2.38 & 11.42 & 12.47 & 12.00 \\ \hline
    0.05 &  0.25 & 2.31 & 11.08 & 12.04 & 11.59 \\ \hline
    0.05 &  0.50 & 2.40 & 11.38 & 12.21 & 11.72 \\ \hline

\multicolumn{6}{c}{GM1nph}\\ \hline
    0 &  0 & 1.99 & 12.39 & 13.81 & 13.60\\ \hline
    0.025 &  0 & 1.84 & 11.91 & 13.26 & 13.19 \\ \hline
    0.05 &  0 & 1.67 & 11.61 & 12.57 & 12.68 \\ \hline
    0.025 &  0.25 & 1.91 & 11.86 & 13.32 & 13.20 \\ \hline
    0.025 &  0.50 & 1.97 & 11.82 & 13.38 & 13.23 \\ \hline
    0.025 &  1 & 2.08 & 11.74 & 13.47 & 13.27 \\ \hline
    0.05 &  0.25 & 1.96 & 11.25 & 12.92 & 12.83 \\ \hline
    0.05 &  0.50 & 2.15 & 11.41 & 13.14 & 12.93 \\ \hline

\end{tabular}
\caption{Some parameters of neutron stars for various EoSs and
$\alpha$ and $\beta$ in models of dark energy with $p_d = -\rho_d
+ \beta \rho_{d}^{2}$ and coupling with matter as $\rho_d = \alpha
\rho_m$. The values of the parameter $\gamma=10$ are in units of
$GM_{\odot}/c^2$.}
\end{center}
\end{table}
From table 2 one can see the same trends as for General Relativity
case. The effect of dark energy is stronger for the GM1nph EoS and
for some values of $\alpha$ and $\beta$ we can reconcile this EoS
with current observational limit of neutron star mass. For example
if $\alpha=0.025$ and $\beta=\beta_{0}$, the maximal mass is
greater by $\sim 0.1 M_{\odot}$ in comparison with pure modified
gravity ($\gamma=10$) or $0.16 M_{\odot}$ greater in comparison
with GR. For $\alpha=0.05$ and $\beta=0.5$ this increasing of the
maximal mass is $0.16 M_{\odot}$ and $0.23 M_{\odot}$
correspondingly. For the relatively stiff MPA1 EoS, even for large
$\beta$ and $\alpha$, the maximal mass is smaller than in $R^2$
gravity while as for APR4 and GM1 EoS, an the increase in the mass
due to dark energy is not so large as for the GM1nph EoS.
\begin{table}
\begin{center}
\begin{tabular}{c|c|c|c|c|c}
\hline
$\alpha$ & $\beta,$ & $M_{max},$ & $R_{max},$ & $R_{1.4},$ & $R_{1.0},$ \\
         & $\beta_{0}$ & $M_{\odot}$ & km & km & km \\ \hline
\multicolumn{6}{c}{GR}\\ \hline
    0 &  0 & 1.88 & 10.50 & 10.68 & 9.88\\ \hline
    0.025 & 0 &  1.79 & 10.27 & 10.57 & 9.76 \\ \hline
    0.050 & 0 &  1.69 & 10.06 & 10.45 & 9.71 \\ \hline
    0.025 & 0.25 & 1.84 & 10.29 & 10.59 & 9.79 \\ \hline
    0.025 & 0.50 & 1.89 & 10.31 & 10.61 & 9.80 \\ \hline
    0.025 & 1   & 1.98 & 10.31 & 10.65 & 9.80 \\ \hline
    0.05 & 0.25 & 1.90 & 10.07 & 10.55 & 9.75 \\ \hline
    0.05 & 0.50 & 2.07 & 10.18 & 10.62 & 9.75 \\ \hline

\multicolumn{6}{c}{$R^2$ gravity with $\gamma=10$}\\ \hline
    0 &  0 & 1.96 & 10.97 & 11.03 & 10.10 \\ \hline
    0.025 & 0 &  1.86 & 10.72 & 10.90 & 10.03 \\ \hline
    0.050 & 0 & 1.74 & 10.50 & 10.78 & 9.94  \\ \hline
    0.025 & 0.25 &  1.92 & 10.81 & 10.94 & 10.04 \\ \hline
    0.025 & 0.50 &  1.98 & 10.68 & 10.96 & 10.02 \\ \hline
    0.025 & 1 &  2.09 & 10.78 & 11.00 & 10.06 \\ \hline
    0.050 & 0.25 & 2.00 & 10.62 & 10.90 & 10.00  \\ \hline
    0.050 & 0.50 & 2.19 & 10.84 & 10.99 & 10.04  \\ \hline

\end{tabular}
\caption{Some parameters of strange stars for various values of
$\alpha$ and $\beta$ in a model of dark energy with $p_d = -\rho_d
+ \beta \rho_{d}^{2}$ and coupling with matter as $\rho_d = \alpha
\rho_m$. }
\end{center}
\end{table}
To conclude this section we need to mention what happens in the
case of strange stars. Its parameters are presented in Table 3 for
GR+DE and $R^2$-gravity+DE models. For the same values of the
parameters as in the neutron stars case, the effect of increasing
mass is more significant. However, unlike neutron stars, the radii
of strange stars with maximal mass and two chosen masses, $1
M_{\odot}$ and $1.4 M_{\odot}$ vary weakly with $\alpha$ and
$\beta$.

\section{Dark energy as a scalar field interacting with gravitational field}

We can also describe dark energy in terms of a scalar-tensor
theory using the equations for the energy density and pressure via
a scalar field $\phi$ with potential $V(\phi)$. In isotropic
cosmology, the scalar field depends only on the cosmic time and,
\begin{equation}
    \rho_d = \epsilon \frac{\dot{\phi}^{2}}{2} + V(\psi),\quad p_d = \epsilon \frac{\dot{\phi}^{2}}{2} - V(\phi). \label{scalar}
\end{equation}
where $\epsilon=\pm 1$ and the minus sign before the kinetic term
corresponds to a phantom field. Using the EoS for  the dark energy
in the form of Eq. (\ref{EoS}), one obtains the following
relations for the scalar field energy density and potential:
\begin{equation}
    \phi(\rho_d) = \phi_{0} + \frac{1}{\sqrt{24\pi}}\int_{\rho_{d0}}^{\rho_{d}}\frac{d\rho_{d}}{\rho_{d}^{1/2}\sqrt{|f(\rho_{d})|}},\label{psi_rho}
\end{equation}
\begin{equation}
    V(\rho_{d}) = \rho_{d} - \frac{f(\rho_d)}{2}.\label{V_rho}
\end{equation}
Here $\phi_0$ is a constant of integration. Combining these
equations we can derive the potential as a function of the scalar
field. For example if the function $f$ is taken as in Eq.
(\ref{EoS_1}), the potential of the scalar field can be written in
explicit form. Using relations,
\begin{equation}
    \phi(\rho_d) = \frac{1}{\sqrt{24\pi |\beta|}}\frac{2}{1-m}\left(\rho_{d}^{\frac{1-m}{2}}-\rho_{d0}^{\frac{1-m}{2}}\right) + \phi_0,\label{psi_rho1}
\end{equation}
\begin{equation}
    V(\rho_{d}) = \rho_{d} - \frac{\beta\rho_{d}^{m}}{2}\label{V_rho1}
\end{equation}
by choosing appropriately the constant $\phi_0$, one obtains
that,
\begin{equation}
    V(\phi) = \left(6\pi|\beta|(1-m)^{2}\phi^{2}\right)^{\frac{1}{1-m}} - \label{Vphi1}
\end{equation}
$$
-\frac{\beta}{2}\left(6\pi|\beta|(1-m)^{2}\phi^2\right)^{\frac{m}{1-m}}.
$$
This result is valid for phantom ($\beta<0$) and quintessence
field ($\beta>0$). It is interesting to consider a cosmological
model with quasi-de Sitter evolution (\ref{EOSDS}). Calculations
yield the following potential as function of scalar field: \be
V(\phi)=\rho_{f} + \lambda\phi^{4}-\mu^{2}\phi^{2}, \label{VDS}
\ee
$$
\lambda=\frac{9\beta^{2}}{16}\rho_f,\quad
\mu^{2}=\frac{3\beta^{2}}{8}\left(\frac{4}{\beta}-1\right)\rho_f.
$$

For stars we can consider this field as part of Lagrangian for
matter, namely,
\begin{equation}
    L_{\phi} =
    -\epsilon\frac{1}{2}g^{\mu\nu}\partial_{\mu}\phi \partial_{\nu}\psi - V(\phi)
    \label{L_scalar}
\end{equation}
where $\epsilon = \pm 1$ corresponds to quintessence or phantom
field correspondingly. For a spherically symmetric star, the
scalar field depends only on the radial coordinate $r$.

We assume that gravitational part of Lagrangian is,
\begin{equation}
S_{g} = -\frac{1}{16\pi}\int d^4 x \sqrt{-g}(1+\eta \phi)R,
\end{equation}
i.e. we consider an interaction between the scalar and
gravitational fields in a simple form. Upon performing a
transformation of metric $\tilde{g}_{\mu\nu}=(1+\eta\phi)
g_{\mu\nu}$, one can write the action in the Einstein frame, \be
S_{g}=\int d^{4} x \sqrt{-\tilde{g}}\left(\frac{\tilde{R}}{16\pi}
-\frac{\tilde{\epsilon}}{2}\tilde{g}^{\mu\nu}\partial_{\mu}\psi\partial_{\nu}\psi-U(\psi)\right),
\ee where redefined the scalar field $\psi$ , which can be
obtained from the relation,
\begin{equation}\label{phi_psi}
\frac{d\phi}{d\psi} = (1+\eta \phi) \left|\epsilon(1+\eta \phi)+\frac{3}{16\pi}\eta^2\right|^{-1/2},
\end{equation}
and the redefined potential in the Einstein frame is
$U(\phi)=(1+\eta \phi)^{-2}V(\phi)$. The parameter
$\tilde{\epsilon}$ is
$$
\tilde{\epsilon} = \mbox{sign}\left(\epsilon(1+\eta \phi)+\frac{3}{16\pi}\eta^2\right).
$$
The metrics in the Einstein and Jordan frames are connected by the
relation,
 \be \label{metric2} d\tilde{s}^{2}=(1+\eta\phi)
ds^{2}=-e^{2\tilde{\nu}}d{t}^{2}+e^{2\tilde{\phi}}{\tilde{dr}}^{2}+\tilde{r}^2d\Omega^2.
\ee Here we write $d\tilde{s}^{2}$ in form equivalent to
(\ref{metr_star}) but with different functions $\tilde{\nu}$ and
$\tilde{\lambda}$. From Eq. (\ref{metric2}) we have that
$\tilde{r}^2=(1+\eta\phi) r^{2}$ and
$e^{2\tilde{\nu}}=(1+\eta\phi) e^{2\nu}$. Combining it with the
following equality,
$$
(1+\eta\phi) e^{2\lambda}dr^{2}=e^{2\tilde{\lambda}}d\tilde{r}^{2}
$$
we obtain that,
$$
e^{-2\lambda}=e^{-2\tilde{\lambda}}\left(1-\frac{1}{2}\eta\tilde{r}A\phi'(\tilde{r})\right)^{2}, \quad A\equiv (1+\eta\phi)^{-1}.
$$
The gravitational mass $m(r)$ is defined in Jordan frame as, \be
m(r) = \frac{r}{2}\left(1-e^{-2\lambda}\right). \ee We can define
the function $\tilde{m}(\tilde{r})$ by using the same relation,
$$
\tilde{m}(\tilde{r}) = \frac{\tilde{r}}{2}\left(1-e^{-2\tilde{\lambda}}\right).
$$
Note that $\tilde{m}(\tilde{r})$ is not the gravitational mass
measured by an observer. But the measured gravitational mass
$m(r)$ can be obtained from a simple relation
$\tilde{m}(\tilde{r})$ as \be m(\tilde{r})=
\frac{\sqrt{A}\tilde{r}}{2}\left(1-\left(1-\frac{2\tilde{m}}{\tilde{r}}\right)\left(1-\frac{1}{2}\eta\tilde{r}A\phi'(\tilde{r})\right)^{2}\right).
\ee
The resulting equations for the metric functions
$\tilde{\lambda}$ and $\tilde{\nu}$ are very similar to the TOV
equations with redefined energy and pressure, and with additional
terms with the energy density and pressure of the scalar field
$\psi$ being:
\be \label{TOV1-1} \frac{1}{4\pi\tilde{r}^2}\frac{d
\tilde{m}}{d\tilde{r}}=
A^{2}\rho+\frac{\tilde{\epsilon}}{2}\left(1-\frac{2\tilde{m}}{\tilde{r}}\right)\left(\frac{d\psi}{d\tilde{r}}\right)^{2}+U(\psi),
\ee

\be \label{TOV2-1}
\frac{1}{4\pi (p+\rho)}\frac{dp}{d\tilde{r}}=-\frac{\tilde{m} + 4\pi
A^{2} p
\tilde{r}^3}{4\pi\tilde{r}(\tilde{r}-2\tilde{m})}-
\frac{\tilde{\epsilon}\tilde{r}}{2}\left(\frac{d\psi}{d\tilde{r}}\right)^{2}+\ee
$$
+\frac{\tilde{r}^2
U(\psi)}{\tilde{r}-2\tilde{m}} +
\frac{\eta}{2}A\frac{d\phi}{d\tilde{r}} ,
$$
The second equation is obtained with using the condition of
hydrodynamic equilibrium,
\begin{equation}\label{hydro-1}
    \frac{dp}{d\tilde{r}}=-(\rho
    +p)\left(\frac{d\tilde{\nu}}{d\tilde{r}}-{\frac{\eta}{2}}A(\phi)\frac{d\phi}{d\tilde{r}}\right).
\end{equation}
Finally, one needs to add the equation of the scalar field
obtained by taking the trace of Einstein equations: \be
\label{TOV3-1}
\tilde{\epsilon}\frac{d\psi}{dr}\triangle_{\tilde{r}}
\psi-\frac{dU}{d\psi}\frac{d\psi}{dr}=-A^3\frac{\eta}{2}\frac{d\phi}{dr}(\rho-3p).
\ee We can rewrite this equation in terms of scalar field $\phi$
by using the relation (\ref{phi_psi}):
\begin{equation}
    \frac{d^2\phi}{d\tilde{r}^2} + \left(\frac{2}{\tilde{r}}+\frac{d\tilde{\nu}}{d\tilde{r}} - \frac{d\tilde{\lambda}}{d\tilde{r}}\right)\frac{d\phi}{d\tilde{r}} + \frac{1}{d\psi/d\phi}\frac{d^{2}\psi}{d\phi^{2}}\left(\frac{d\phi}{d\tilde{r}}\right)^{2}-
\end{equation}
$$
-e^{2\tilde{\lambda}}\frac{\tilde{\epsilon}}{(d\psi/d\phi)^{2}}\frac{dU}{d\phi} = -\frac{\tilde{\epsilon}A^{3}}{(d\psi/d\phi)^{2}}\frac{\eta}{2}e^{2\tilde{\lambda}}(\rho - 3p).
$$
Equations (\ref{TOV1-1}), (\ref{TOV2-1}) with (\ref{TOV3-1}) can
be integrated numerically for the following initial conditions at
the center of star:
$$
\theta(0)=1,\quad \tilde{\mu}(0)=0, \quad \phi(0)=\phi_{0}, \quad \frac{d\phi(0)}{d\tilde{r}}=0.
$$
The condition of asymptotic flatness requires that
$$\phi\rightarrow 0 \quad \mbox{at} \quad r\rightarrow \infty. $$
It is convenient to analyze the system of equations in the
Einstein frame and then after calculations transform the
corresponding quantities back to the Jordan frame.

Our analysis shows that for the potential (\ref{Vphi1})
($\beta>0$) and $m=1/2$, the solutions of TOV equations do not
yield results which can be interpreted from physical viewpoint.
The model with (\ref{VDS})  is more interesting. If the current
dark energy density in the Universe is close to $\rho_f$, the
value of $\beta\rho_f$ can be sufficiently large in comparison
with $\rho_d$, and therefore the parameter $\beta$ can be very
large. Of course, since $\rho_f\sim 10^{-27}$ g/cm$^3$ and this
value is very small in comparison with densities in neutron stars,
the value of the parameter $\beta$ should be vary large in order
to have some observable effects. For consistency with astronomical
observations, this requires that the quantity $\rho_d/\rho_f$
should be close to unity with incredible precision. Although it
looks as an extremely fine-tuned scenario, we consider such a
model as an illustrative example. For $\beta>>1$ our potential is
$$
V(\phi) = \rho_f + \frac{9\beta^2}{16}\rho_f \phi^4 + \frac{3\beta^2}{8}\rho_f \phi^2.
$$
If we introduce parameter,
$$
\rho_0 = M_{\odot} / r_{g}^3, \quad r_g = GM_{\odot}/c^2
$$
we can see some effects from dark energy if,
$$
\beta^2 \rho_{f}/\rho_{0} \approx 0.01.
$$
\begin{figure}
    \centering
    \includegraphics[scale=0.37]{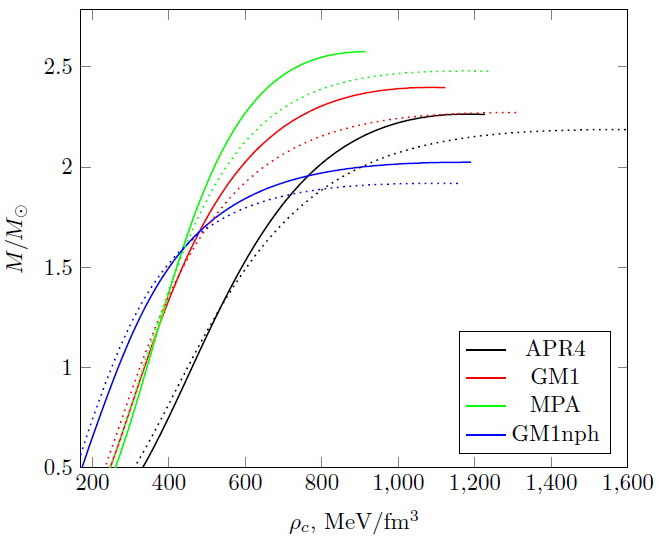}\includegraphics[scale=0.37]{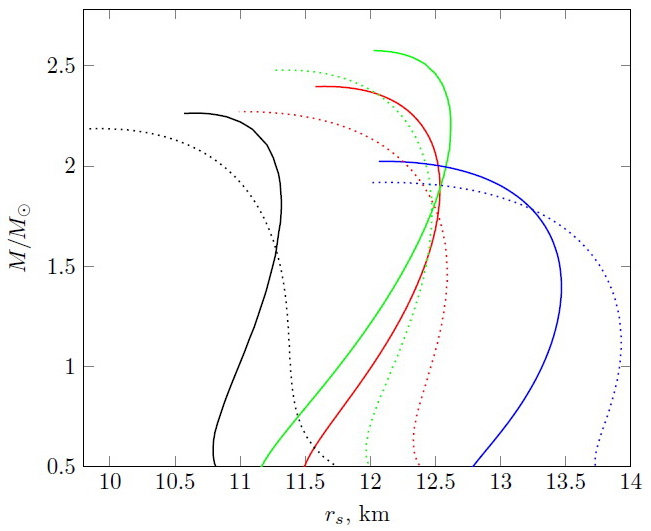}
    \caption{The dependence of the stellar mass as a function of the central density (left panel)
and mass-radius relation (right panel) in a dark energy model with
potential (\ref{VDS}) and non-minimal interaction with gravity
(parameter $\eta=1$) for neutron stars.  The parameter $\beta$ is
chosen so that $\beta^2\rho_f/\rho_0=0.01$. Dotted lines
correspond to GR.}
    \label{fig13}
\end{figure}
This corresponds to $\beta\approx 10^{21}$, which is remarkable.
Of course it seems very unrealistic. It is interesting to note
that this model of scalar field  gives results very similar to
$R^2$ gravity. For illustration we take $\eta = 1$. The results of
our numerical analysis are given in Fig. 13 for neutron stars and
in Table 4.

\section{Concluding Remarks}

We considered the possible effects of dark energy on the
parameters of relativistic compact stars, such as the
gravitational mass and radius. Our analysis involved realistic
EoSs for stellar matter in neutron stars and hypothetical quark
stars. The simple assumption about possible interaction between
dark energy and matter is imposed from the beginning of our
analysis. For dark energy EoSs we investigated the realistic model
satisfying the current astrophysical observational data.

For $\Lambda$-like model of dark energy, the maximal mass of the
star for given EoS decreases due to negative contribution to total
pressure. For dark energy with $w>-1$ (quintessence) it is
possible to construct models in which for large densities, the
contribution to pressure is positive and the effective EoS for
stellar matter and dark energy became stronger. As a result the
maximal mass increases and the corresponding radius decreases.
Models of dark energy with $w<-1$ and a direct coupling of dark
energy with matter seem unrealistic because the effect of EoS
softening will intensify even more, and most of the EoSs will be
ruled out by observational limits on neutron star mass.

We investigated the combined effect of dark energy as ideal fluid
interacting with matter and an $R^2$-term in the action for the
gravitational field. In this case, the effect of increasing mass
for neutron stars due to $R^2$-term is partially eliminated by
dark energy with $p_d\approx -\rho_d$ for $\alpha\approx 0.025$.
This result is independent from the EoS of the stellar matter.
Another interesting feature is that the radii for neutron stars
with small masses are considerably reduced in comparison with GR.
For dark energy with $p_d = -\rho_d + \beta\rho_d^2$, the effects
from a possible dark energy component and $R^2$-term reinforce
each other. For soft EoS with hyperons, the combined effects of
dark energy and $R^2$ gravity are stronger and for some parameters
we can reconcile therefore such EoS with results of current
astronomical observations. An interesting feature occurs for
strange stars. For the same parameters as in a case of neutron
stars, the effect of increasing mass is more pronounced. However,
unlike neutron stars, the radii of strange stars with maximal mass
vary weakly with the parameters $\alpha$ and $\beta$. Another way
to describe dark energy is by using a scalar field theory and
consider non-minimal interaction between gravity and the scalar
field. For dark energy models compatible with astronomical
observations, this approach does not always give results
acceptable from a physical point of view. Simple dark energy
models with asymptotic de Sitter expansion, correspond to
power-law potential of the scalar field and give results very
similar to $R^2$ gravity. This correspondence is not unexpected
because $R^2$ gravity as is well known can be described in frames
of equivalent scalar field theory with power-law potentials.

In conclusion one needs to stress that we cannot say that the EoS
for dense stellar matter is known accurately. Therefore possible
effects of dark energy and modified gravity can be hidden, by lack
of knowledge about the exact EoS of nuclear matter. The major task
is to discriminate the possible dark energy effects in neutron
stars from EoS variations.

\section*{Acknowledgments}

This work was partially supported by MICINN (Spain), project
PID2019-104397GB-I00  and by the program Unidad de Excelencia
Maria de Maeztu CEX2020-001058-M, Spain (SDO). This work was
supported by Ministry of Education and Science (Russia), project
075-02-2021-1748 (AVA).

\section*{Data availability}
No new data were generated or analysed
in support of this research.

\label{lastpage}


\begin{thebibliography}{99}

\bibitem{Riess1} Riess A.G. {et al.}  [Supernova Search Team Collaboration], 1998, AJ, {116}, 1009 [arXiv:astro-ph/9805201]

\bibitem{Riess2} Riess A.G. {et al.}  [Supernova Search Team Collaboration], 2004, ApJ,  {607}, 665 [arXiv:astro-ph/0402512]

\bibitem{Perlmutter} Perlmutter S. {et al.}  [Supernova Cosmology Project Collaboration], 1999, ApJ, {517}, 565 [arXiv:astro-ph/9812133]

\bibitem{Dark-6} M.~Li, X.~Li, S.~Wang and Y.~Wang, Commun.\ Theor.\ Phys. {\bf 56}, 525 (2011)

\bibitem{Ca} K. Bamba, S. Capozziello, S. Nojiri and S.D. Odintsov, Astr. Space Sci {\bf 342}, 155 (2012) [arXiv:1205.3421].

\bibitem{Dark-7} C. Clarkson, G. Ellis, J. Larena and O. Umeh, Rept. Prog. Phys. {\bf 74}, 112901 (2011).

\bibitem{Kowalski} M.~Kowalski, Ap.\ J.\ {\bf 686}, 74 (2008).

\bibitem{Weinberg} Weinberg S., 1989, Rev. Mod. Phys., {61}, 1

\bibitem{reviews1}
 Capozziello S., De Laurentis M., 2011,
   Phys.\ Rept.,  {509}, 167
   [arXiv:1108.6266 [gr-qc]].

\bibitem{reviews2}
Capozziello S., Faraoni V.
 \textit{Beyond Einstein Gravity : A Survey of Gravitational Theories for Cosmology and Astrophysics}, 2011,
  Fundam.\ Theor.\ Phys.,  {170}, Springer, Dordrecht


\bibitem{reviews4}
 Nojiri S., Odintsov S.D., Oikonomou V.K., 2017,
  Phys.\ Rept., {692}, 1
  [arXiv:1705.11098 [gr-qc]]

\bibitem{book}
Nojiri S., Odintsov S.D., 2011,
   Phys.\ Rept.,  {505}, 59

\bibitem{reviews5}
de la Cruz-Dombriz A., Saez-Gomez D., 2012,
  Entrp, {14}, 1717
  [arXiv:1207.2663 [gr-qc]].

\bibitem{reviews6}
Olmo G.J., 2011,
  IJMPD, {20}, 413
  [arXiv:1101.3864 [gr-qc]].

\bibitem{dimo} Dimopoulos K., 2021, \textit{Introduction to Cosmic Inflation and Dark Energy}, CRC Press


\bibitem{Nojiri:2003ft}
Nojiri S., Odintsov S.D., 2003,
PhRvD, {68}, 123512 [arXiv:hep-th/0307288]
  

\bibitem{Astashenok:2020qds}
Astashenok A.V., Capozziello S., Odintsov S.D.,
Oikonomou V.K., 2020,
Phys. Lett. B, {811}, 135910
[arXiv:2008.10884 [gr-qc]]

\bibitem{Astashenok:2021peo}
Astashenok A.V., Capozziello S., Odintsov S.D.,
Oikonomou V.K., 2021,
 Phys. Lett.  B, {816}, 136222
 [arXiv:2103.04144 [gr-qc]]

\bibitem{Capozziello:2015yza}
Capozziello S.,, De Laurentis M., Farinelli R., Odintsov S.D., 2016,
Phys. Rev. D, {93}, 023501
[arXiv:1509.04163 [gr-qc]]

\bibitem{Astashenok:2014nua}
Astashenok A.V., Capozziello S., Odintsov S.D., 2015,
JCAP, 01, 001 
[arXiv:1408.3856 [gr-qc]]

\bibitem{Astashenok:2020cfv} Astashenok A.V., Odintsov S.D., 2020,
MNRAS, {493}, 78.

\bibitem{Arapoglu:2010rz}
Arapoglu A.S., Deliduman C., Eksi K.Y., 2011,
JCAP, {07}, 020 
[arXiv:1003.3179 [gr-qc]]

\bibitem{Panotopoulos:2021sbf}
Panotopoulos G.,, Tangphati T., Banerjee A., Jasim M.K.,
[arXiv:2104.00590 [gr-qc]]

\bibitem{Lobato:2020fxt}
Lobato R., Louren\c{c}o O., Moraes P.H.R.S., Lenzi C.H., de
Avellar M., de Paula W., Dutra M., Malheiro M., 2020,
JCAP, {12}, 039 
[arXiv:2009.04696 [astro-ph.HE]]

\bibitem{Oikonomou:2021iid}
Oikonomou V.K., 2021,
Class. Quant. Grav., {38}, 175005


\bibitem{Odintsov:2021qbq}
Odintsov S.D., Oikonomou V.K., 2021,
Phys. Dark Univ., {32}, 100805
[arXiv:2103.07725 [gr-qc]]

\bibitem{Wojnar3} Olmo G.J., Rubiera-Garcia D., Wojnar A., 2020, Phys. Rept., 876, 1 [arXiv:1912.05202 [gr-qc]]

\bibitem{Pani:2014jra}
Pani P., Berti E., 2014,
Phys. Rev. D, {90}, 024025
[arXiv:1405.4547 [gr-qc]]

\bibitem{Doneva:2013qva}
Doneva D.D., Yazadjiev S.S., Stergioulas N., Kokkotas K.D., 2013,
Phys. Rev. D, {88}, 084060
[arXiv:1309.0605 [gr-qc]]

\bibitem{Horbatsch:2015bua}
Horbatsch M., Silva H.O., Gerosa D., Pani P., Berti E.,
Gualtieri L., Sperhake U., 2015,
Class. Quant. Grav., {32}, 204001
[arXiv:1505.07462 [gr-qc]]

\bibitem{Silva:2014fca}
Silva H.O., Macedo C.F.B., Berti E., Crispino L.C.B., 2015,
Class. Quant. Grav., {32}, 145008
[arXiv:1411.6286 [gr-qc]]

\bibitem{Chew:2019lsa}
Chew X.Y., Kleihaus B.,
Kunz J., Dzhunushaliev V., Folomeev V., 2019,
Phys. Rev. D, {100}, 044019
[arXiv:1906.08742 [gr-qc]]

\bibitem{Blazquez-Salcedo:2020ibb}
Bl\'azquez-Salcedo J.L., F.~Scen Khoo and J.~Kunz,
EPL \textbf{130} (2020) no.5, 50002
[arXiv:2001.09117 [gr-qc]].

\bibitem{Motahar:2017blm}
Motahar Z., Bl\'azquez-Salcedo J.L., Kleihaus B.,
Kunz J., 2017,
Phys. Rev. D, {96}, 064046
[arXiv:1707.05280 [gr-qc]]

\bibitem{NOT} S.~Nojiri, S.~D.~Odintsov and S.~Tsujikawa, Phys.\ Rev.\ D {\bf 71}, 063004 (2005) [hep-th/0501025]

\bibitem{AP4} Akmal A. and Pandharipande V. R. and Ravenhall, D. G., Phys. Rev. C 58, 1804 (1998).

\bibitem{MPA} Muther, H. and Prakash, M. and Ainsworth, T. L., Phys. Lett. B199, 469 (1987).

\bibitem{GM} Glendenning, N. K. and Moszkowski, S. A., Phys. Rev.
Lett. 67, 2414 (1991).

\bibitem{GMnph} Oertel, M. and Providencia, C. and Gulminelli, F. and
Raduta, A. R., J. Phys. G Nucl. Phys. 42, 075202 (2015).

\bibitem{QS} E. Witten, Phys. Rev. D 30 (1984) 272.

\bibitem{QS1} N. Itoh, Progress of Theoretical Physics 44, 291 (1970).

\bibitem{MIT} R.L. Jaffe, F. E. Low, Phys. Rev. D. 19, 2105 (1979).

\bibitem{MIT1} Yu. Simonov, Phys. Lett. B. 107 (1981).

\bibitem{MIT2} N. Stergioulas, Living Rev. Rel. 6, 3 (2003) arXiv:gr-qc/0302034 [gr-qc].

\bibitem{Antoniadis} Antoniadis J. et al., 2013, Science, 340 (6131), 1233232

\bibitem{Demorest} Demorest P.B. et al., 2010,  Nature, 467 (7319), 1081

\bibitem{Romani} Romani R.W. et al., 2022, AJL, 934, 6

































\end{thebibliography}
\end{document}